\documentclass[twocolumn,english,aps,prl,showpacs]{revtex4-1}
\usepackage[T1]{fontenc}
\usepackage[latin9]{inputenc}
\usepackage{verbatim}
\usepackage{graphicx}

\makeatletter

\providecommand{\tabularnewline}{\\}

\@ifundefined{textcolor}{}
{%
 \definecolor{BLACK}{gray}{0}
 \definecolor{WHITE}{gray}{1}
 \definecolor{RED}{rgb}{1,0,0}
 \definecolor{GREEN}{rgb}{0,1,0}
 \definecolor{BLUE}{rgb}{0,0,1}
 \definecolor{CYAN}{cmyk}{1,0,0,0}
 \definecolor{MAGENTA}{cmyk}{0,1,0,0}
 \definecolor{YELLOW}{cmyk}{0,0,1,0}
 }

\newcommand{\be}{\begin{equation}}
\newcommand{\ee}{\end{equation}}
\newcommand{\bea}{\begin{eqnarray}}
\newcommand{\eea}{\end{eqnarray}}

\newcommand{\BE}{\begin{eqnarray}}
\newcommand{\EE}{\end{eqnarray}}
\newcommand{\BEn}{\begin{eqnarray*}}
\newcommand{\EEn}{\end{eqnarray*}}
\newcommand{\barr}{\begin{array}}
\newcommand{\earr}{\end{array}}

\newcommand{\bit}{\begin{itemize}}
\newcommand{\eit}{\end{itemize}}
\newcommand{\bc}{\begin{center}}
\newcommand{\ec}{\end{center}}
\newcommand{\ben}{\begin{enumerate}}
\newcommand{\een}{\end{enumerate}}

\makeatother

\usepackage{babel}

\begin{document}

\title{Prediction accuracy and sloppiness of log-periodic functions }

\author{David Brée}

\affiliation{School of Computer Science, University of Manchester, Machester M13
9PL, U.K.}

\affiliation{previously at ISI Foundation, Vialle Settimio Severo 65, 14011 Turin,
Italy}

\email{dbree@cs.man.ac.uk}

\author{Damien Challet}

\affiliation{Physics Department, University of Fribourg, ch. du Musée 2, 1700
Fribourg, Switzerland}

\affiliation{previously at ISI Foundation, Vialle Settimio Severo 65, 14011 Turin,
Italy}

\email{damien.challet@unifr.ch}

\author{Pier Paolo Peirano}

\affiliation{CFM, 6, bvd Haussmann, 75009 Paris, France}

\affiliation{previously at ISI Foundation, Vialle Settimio Severo 65, 14011 Turin,
Italy}

\email{ppp13t@gmail.com}

\date{\today}
\begin{abstract}
We show that log-periodic power-law (LPPL) functions are intrinsically
very hard to fit to time series. This comes from their sloppiness,
the squared residuals depending very much on some combinations of
parameters and very little on other ones. The time of singularity
that is supposed to give an estimate of the day of the crash belongs
to the latter category. We discuss in detail why and how the fitting
procedure must take into account the sloppy nature of this kind of
model. We then test the reliability of LPPLs on synthetic AR(1) data
replicating the Hang Seng 1987 crash and show that even this case
is borderline regarding predictability of divergence time. We finally
argue that current methods used to estimate a probabilistic time window
for the divergence time are likely to be over-optimistic.
\end{abstract}
\maketitle

\section{Introduction}

Log-periodic functions have received much attention because of the
claim that they could be used to predict the times of singularities.
While they are known to occur in hierarchical discrete scale-free
networks \cite{SH:2005}, they have been claimed to have been observed
in many types of natural time/size series: earthquakes \cite{NTG:1995,B:1997,T:2002,CKYOSTS:2008},
icequakes \cite{FFS:2009}, forest fires \cite{MMT:2005} as well
as evolutionary trees \cite{NCG:2000}, although such claims have
not gone unchallenged \cite{HSS:2000}. But the most noticed application
of such functions is to speculative bubbles of stock indices \cite{SJ:1997,JLS:2000,SJ:2001,BCSWZ:2009,sornette1996stock},
foreign exchange rates \cite{MGFD:2005}, real estate \cite{ZS:2006}
and commodity prices \cite{DKO:2008,DKOS:2008} as well as downward
spirals during the burst of the bubble \cite{JS:2001,LM:2004}. Given
the importance of such phenomena, and the possibly important consequences
of finding a universal model that could be applied to this remarkable
variety of bubbles, it is of course necessary to assess the statistical
signifiance of LPPLs regarding crashes, i.e., the predictive power
of log-periodic functions in this context. The question is still unsettled
as of yet \cite{LPC:1999,F:2001,B:2003,CF:2006,SZ:2006,GFPR:2008}.
Problems are indeed numerous: what definition of a bubble and a crash
to adopt\cite{J:2009,LRS:2009}%
\footnote{For example \cite{LRS:2009} recently claim that the LPPL model has
few false positives on the grounds that all five of the bubbles that
they identified on the S\&P500 between 3 January 1950 and 21 November
2008 {}``preceded well-know crashes.'' Using a criterion for a crash
of a drop by > 15\% within 12 weeks identifies 12 crashes in this
period, but only two of these (25/8/1987, 17/7/1998) were among the
five claimed by \cite[Table 3]{LRS:2009}; the other three {}``well-known
crashes'' (in 1994, 1997 and 1999) do not meet these quite lenient
crash criteria.%
}? should the price in a bubble always be increasing \cite{BM:2003}?,
should one impose contraints on the fitted parameters \cite{JS:2001}?
where to start a fit of a bubble%
\footnote{While the beginning of a bubble is supposed to be given by the lowest
value since the previous crash, it may be moved from this minimum
to some later date; in \cite{JS:2001} the beginnings of half of the
eight bubbles on the Hang Seng up to 1998 were thus moved \cite{BJ:2010}. %
}? what test of goodness of fit to use%
\footnote{Lomb periodograms \cite{B:2003} or the residuals \cite{GFPR:2008,LRS:2009}
have been proposed, with mixed results. %
}? why having different lengths of the data window greatly affects
the parameters of the best fit of the LPPL to the data \cite{F:2001}?
why leaving out a few data points can alter the parameters of the
best fit sufficiently to change a no/bubble decision (see e.g. \cite[footnote 4]{LRS:2009})?
why is the fitting error very sensitive to small (but not large!)
changes in one of the parameters of the model \cite{BJ:2010}?

What contributes to most if not all of these difficulties is that
a stable best fit of an LPPL to the data is very hard to determine.
Here we aim to show that this comes from the fact LPPLs belong to
the family of sloppy functions, a terminology introduced in a series
of papers by Sethna \textit{et al} \cite{brown2003statistical,frederiksen2004bayesian,SethnaSloppyPRL,gutenkunst2007universally};
we will discuss in details what this means when applying LPPLs to
noisy time series.

\section{Sloppiness}

Let us denote by $p(t)$ the time series to be fitted, $f$ the fitting
function and $\Phi$ the set of parameters. Least-squares fits minimise
$S=\sum_{t=t_{0}}^{t_{1}}[f_{\Phi}(t)-p(t)]^{2}/(t_{1}-t_{0}-n)$,
where $t_{1}<t_{c}$, the time when the singularity (crash) occurs,
and $n$ is the number of free parameters. The best fit $\hat{\Phi}$
to some given data corresponds by definition to the minimum of $S$,
therefore, close to $\hat{\Phi}$, $S\simeq S_{0}+\sum_{x,y\in P}\left.\frac{\partial^{2}S}{\partial x\partial y}\right|_{\Phi=\hat{\Phi}}(x-\hat{x})(y-\hat{y})$.
Assuming that $\hat{\Phi}$ does not sit on a boundary of the parameter
space, the curvature of $S$ in a neighborhood of $\hat{\Phi}$ is
positive; as a consequence, the Hessian $\frac{\partial^{2}S}{\partial x\partial y}|_{\Phi=\hat{\Phi}}$
is positive-definite, thus all its eigenvalues are positive.

\begin{table*}
\begin{tabular}{|c|c|c|c|c|c|c|c|}
\hline 
$\lambda$ & $A$ & $B$ & $C$ & $t_{c}$ & $\alpha$ & $\omega$ & $\phi$\tabularnewline
\hline
\hline 
1.11$\cdot$10$^{9}$ & -0.00004 & -0.00022 & 0.03198 & 0.00012 & \textbf{0.99943} & 0.01085 & 0.00158\tabularnewline
\hline 
1.21$\cdot$10$^{7}$ & 0.00001 & 0.00005 & \textbf{0.99949} & -0.00016 & -0.03200 & 0.00176 & 0.00060\tabularnewline
\hline 
1.49$\cdot$10$^{6}$ & -0.00009 & -0.00018 & -0.00219 & -0.00146 & -0.01090 & \textbf{-0.98737} & \textbf{-0.15804}\tabularnewline
\hline 
2.89$\cdot$10$^{2}$ & 0.00862 & 0.01599 & 0.00025 & \textbf{-0.27339} & -0.00012 & \textbf{0.15239} & \textbf{-0.94958}\tabularnewline
\hline 
3.13$\cdot$10$^{1}$ & 0.03973 & 0.08183 & -0.00024 & \textbf{-0.95699} & 0.00017 & -0.04192 & \textbf{0.27063}\tabularnewline
\hline 
2.48$\cdot$10$^{-1}$ & \textbf{0.39153} & \textbf{0.91509} & -0.00003 & 0.09615 & 0.00020 & 0.00102 & -0.00855\tabularnewline
\hline 
6.50$\cdot$10$^{-3}$ & \textbf{-0.91915} & \textbf{0.39367} & -0.00001 & -0.01353 & 0.00005 & -0.00031 & 0.00213\tabularnewline
\hline
\end{tabular}\caption{Eigenvalues, $\lambda$, and associated eigenvectors of the best fit
of real price for the 1987 crash. Components of absolute value larger
than 0.1 are in bold face. \label{tab:1987-eigen}}

\end{table*}

\begin{table*}
\begin{tabular}{|c|c|c|c|c|c|c|c|}
\hline 
$\lambda$ & $A$ & $B$ & $C$ & $t_{c}$ & $\alpha$ & $\omega$ & $\phi$\tabularnewline
\hline
\hline 
4.16$\cdot$10$^{4}$ & 0.00621 & \textbf{0.99935} & -0.00141 & -0.00001 & -0.03546 & -0.00241 & -0.00035\tabularnewline
\hline 
1.12$\cdot$10$^{0}$ & \textbf{0.38694} & 0.01331 & 0.01048 & 0.00036 & \textbf{0.38499} & \textbf{0.82776} & \textbf{0.12877}\tabularnewline
\hline 
5.12$\cdot$10$^{-1}$ & 0.07433 & 0.00942 & \textbf{0.95117} & -0.00080 & \textbf{0.25177} & \textbf{-0.16023} & -0.02445\tabularnewline
\hline 
3.56$\cdot$10$^{-1}$ & \textbf{-0.73357} & -0.00726 & \textbf{0.24417} & 0.00276 & -\textbf{0.37764} & \textbf{0.50250} & 0.08409\tabularnewline
\hline 
1.03$\cdot$10$^{-1}$ & \textbf{-0.55369} & 0.03139 & \textbf{-0.18851} & 0.0080 & \textbf{0.80280} & \textbf{-0.10934} & -0.02146\tabularnewline
\hline 
4.48$\cdot$10$^{-5}$ & -0.00226 & -0.00015 & 0.00260 & \textbf{-0.20087} & -0.00560 & \textbf{0.15421} & \textbf{-0.96738}\tabularnewline
\hline 
1.86$\cdot$10$^{-5}$ & 0.00198 & -0.00002 & 0.00078 & \textbf{0.97961} & -0.00067 & 0.02986 & \textbf{-0.19864}\tabularnewline
\hline
\end{tabular}\caption{Eigenvalues, $\lambda$, and associated eigenvectors of the best fit
of log price for the 1987 crash. Components of absolute value larger
than 0.1 are in bold face. \label{tab:1987-eigen-log}}

\end{table*}

As shown recently in a series of papers (e.g. \cite{brown2003statistical,frederiksen2004bayesian,SethnaSloppyPRL,gutenkunst2007universally}),
sloppy models are characterized by a separation of Hessian eigenvalues
by orders of magnitude, which is all the more likely and evident when
the models have many parameters. Combinations of parameters corresponding
to larger eigenvalues are called stiff, while those corresponding
to small eigenvalues are called sloppy. In other words, varying slightly
a stiff parameter combination has a large influence on $S$, while
changing sloppy combinations of parameters does not modify substantially
$S$. This has two consequences, discussed in detail in the following
sections: first, fitting sloppy functions must be done carefully;
second, out of sample predictions from the best-fit values of sloppy
parameters may be imprecise: as the noise from sample to sample changes,
the fitted values of sloppy parameters are likely to change greatly. 

Let us apply this reasoning to the fitting of financial index prices
$p(t)$ with log-periodic functions, as used originally in this context
in \cite{sornette1996stock}, \[
f_{LP}(t)=A+B(t_{c}-t)^{\alpha}[1+C\cos(\log(t_{c}-t)+\phi))].\]

This is a seven-parameter fit, but as already noted in the original
paper, minimizing $S$ with respect to $A$, \textbf{$B$}, and $C$
yields linear equations, which reduces the non-linear part of the
fitting problem to four parameters. However, sloppiness concerns a
priori all seven parameters. This is why we shall keep them all, i.e.
$\hat{\Phi}=\{\hat{A},\hat{B},\hat{t}_{c},\hat{\alpha},\hat{C},\hat{\omega},\hat{\phi}\}$,
in order to give a fuller account of sloppiness. Once we understand
what respective importance $A$, \textbf{$B$}, and $C$ have in $S$,
we will be able to focus on the other parameters. 

It turns out that log-periodic functions are very sloppy: every crash
we fitted resulted in a clear separation of eigenvalues by orders
of magnitude. Let us take for example the 1987 crash in the Hang Seng
index. The eigenvalues and eigenvectors of the Hessian of the best
fit obtained by using the Levenberg-Marquardt algorithm to fit the
834 days preceding the crash, and retaining the best of a set of 20000
initial conditions, are shown for the best fit to real prices in Table
\ref{tab:1987-eigen} and to log prices in Table \ref{tab:1987-eigen-log}. 

These tables contain several relevant pieces of information. First,
the largest eigenvalue is at least 9 orders of magnitude larger than
the smallest one, a definite signature of sloppiness; in addition,
the eigenvalues are well spread over these orders of magnitudes. The
associated eigenvectors confirm the wisdom that the stiffer a direction,
the more likely that it is close to an axis, and reveresely for sloppy
eigenvalues\cite{gutenkunst2007universally}. Next, the eigenvectors
vary from crash to crash and can be quite different between real and
log-prices: for the 1987 crash, the linear-fit parameters ($A$, $B$,
and $C$) are completely disconnected from the other ones only in
the case of real prices; curiously, this is not systematic, as both
log and real prices of the 1997 crash lead to disconnected eigenvectors,
for instance. When the eigenvectors associted to $A$, $B$, and $C$
are not completely disconnected from the other four parameters, one
should not fit them separately when estimating the error associated
with $t_{c}$, for instance (see section Discussion below and \cite{brown2003statistical,frederiksen2004bayesian}).

\begin{figure*}
\includegraphics[scale=0.4]{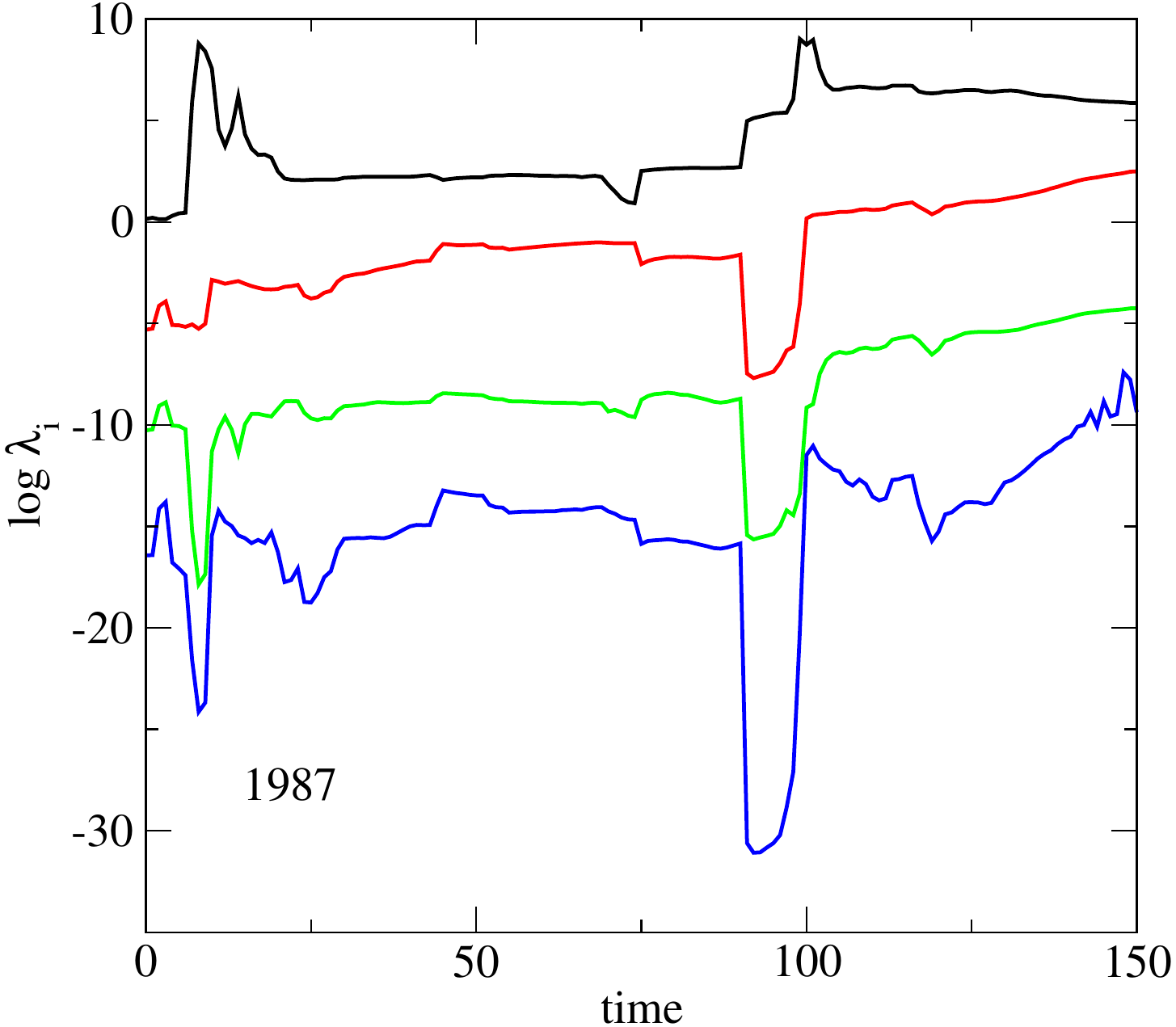}\includegraphics[scale=0.4]{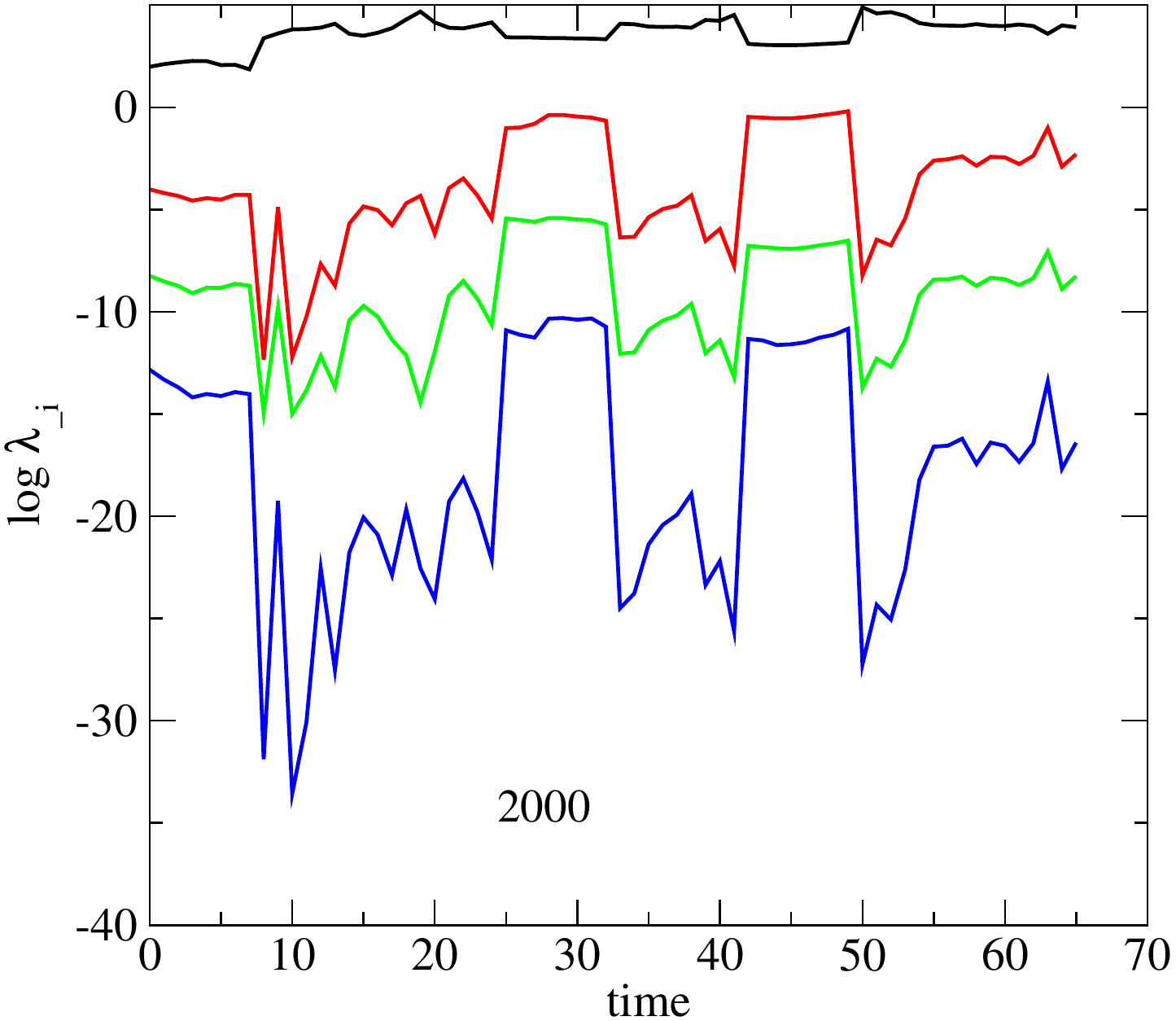}\includegraphics[scale=0.4]{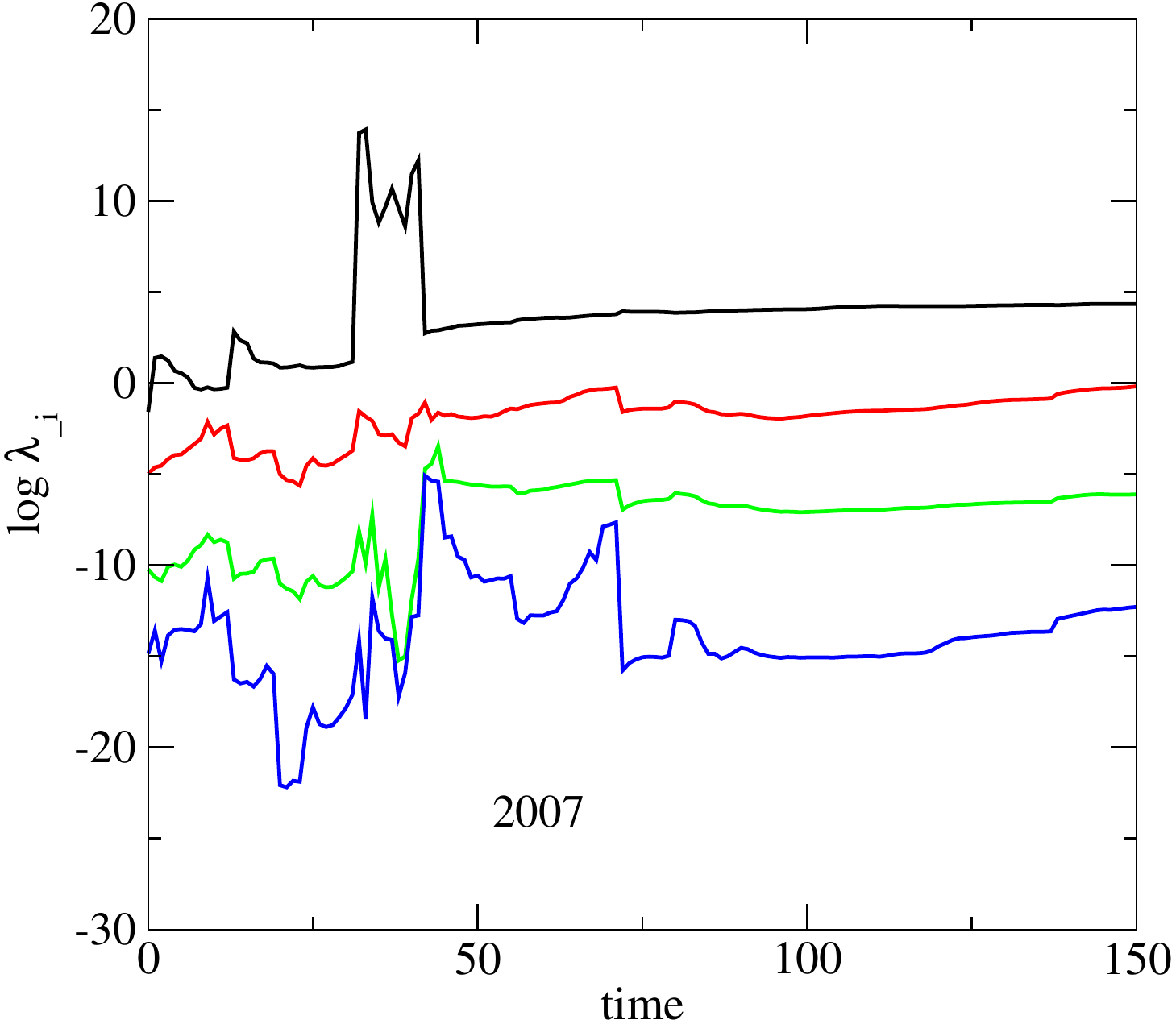}

\caption{Eigenvalues associated to the four parameters requiring non-linear
fitting as a function of time in days immediately before $t_{c}$
for the 1987, 2000, and 2007 crashes (log prices).\label{fig:Eigenvalues-associated-to}}

\end{figure*}

We are of course chiefly interested in the role of $t_{c}$: this
crucial parameter turns out to be one of the most sloppy parameters
and, as a consequence, its associated eigenvector is not along the
$t_{c}$ axis but also comprises the phase $\phi$ and the frequency
$\omega$, meaning that in order to fit $t_{c}$ precisely, one should
take this inter-dependency into account properly, which is not the
case in the state-of-the-art papers on the topic that all rely on
Levenberg-Marquart algorithm (see below for remedies). We also note
that it seems slightly less sloppy for real prices than for log prices
for the 1987 crash.

Sloppiness is intrisic to the LPPL equation, not only to the 1987
crash, nor just to dangerous times just before a crash. In order to
convince oneself of this important point, Figure \ref{fig:Eigenvalues-associated-to}
plots the four eigenvalues associated with the parameters requiring
a non-linear fit as a function of time in the 150 days preceding the
1987, 2000, and 2007 crashes on the Hang Seng, chosen randomly; it
is obvious that the eigenvalues are well-spaced and that their typical
spacing stays very large in the whole time series; their structure
is also constant, with no crossing of eigenvalues. We have found the
same behaviour for all the crashes investigated. Note, however, that
some crashes lead to more sloppy fits than others, i.e., with an even
larger eigenvalue separation.

It should be noted that in principle, some sloppy models can be unsloppied
by a suitable change of fitting functions. For instance, fitting a
function in $[0,1]$ with a sum of exponentials is known to be ill-posed
\cite{BosSumExp}. However, using Hermite's polynoms lifts the sloppiness
of exponentials \cite{SethnaSloppyPRL}. Unfortunately, this approach
relies on a symmetry assumption between the parameters that does not
hold for LPPL. 

Sloppiness has important consequences and, despite its negative connotation,
these are not only negative. However, being aware that LPPLs are sloppy
models helps understand several important aspects of making predictions
with an LPPL, in particular with respect to the uncertainty associated
to the most sloppy parameters; this will be discussed in the next
few sections.

\section{Consequences of Sloppiness}

\subsection{Sensitivity of $t_{c}$}

The main result of the previous section is that not only are LPPL
functions sloppy, but that varying $t_{c}$ together with $\phi$
has little influence on square residuals. Reversely, changing slightly
the input will vary tremendously $t_{c}$. This explains first why
the diagnostic of a bubble is sometimes sensitive to the addition
or deletion of a single data point. By extension, the sensitivity
of $t_{c}$ to noise must be investigated and one must understand
how reliable can the fits of LPPL to noisy data be.

Quite tellingly, early papers using LPPL to predict various kinds
of crashes used only a single fit, which, of course, is problematic
in the light of sloppiness. Recent papers try to build a probabilistic
window for $t_{c}$ \cite{sornette1996stock,BCSWZ:2009,YWS10,JZSWBC:2010}.
The problem one faces is to estimate a probability distribution for
$t_{c}$ from a \emph{single} noisy time series. The methods consists
essentially in varying the beginning and end of the time series, thereby
obtaining a distribution of fitted values for $t_{c}.$ But this only
happens because LPPLs are sloppy and because $t_{c}$ is one of the
least relevant variables in the fit. Thus, this new method uses the
intrisic imprecision of LPPL regarding $t_{c}$. This is the positive
side of sloppiness. The negative side is of course that the imprecision
on $t_{c}$ is a priori very large. In addition, there is no real
guarantee that the distribution of $t_{c}$ thus obtained corresponds
to anything meaningful. As we shall explain below, special methods
have been devised for sloppy functions that are able to give reliable
probability distributions for fitted variables from a single time
series.

\subsection{Fitting LPPL }

First, using simple fitting algorithms is bound to be problematic
for sloppy functions (see e.g. the discussion in \cite{brown2003statistical})
as most of them approximate to a first order the cost function variation
when trying to find the next move in the parameter space. In the case
of sloppy functions, however, one needs to take into account not only
the gradient, but also the curvature of the cost landscape by computing
the eigenvectors and following them, which is computationally more
costly. A computational compromise is the Levenberg-Marquart method
(used by people studying LPPL ever since the original paper) which
approximates the Hessian with a product of gradients, thus implicitely
assuming that the eigenvectors do not deviate much from the axes.
While this is a reasonable approximation as regards some eigenvalues,
as seen in Tables \ref{tab:1987-eigen} and \ref{tab:1987-eigen-log},
it breaks down in particular for $t_{c}$: this means that reaching
a correct estimate of $t_{c}$ requires more sophisticated methods,
such as the Rosebroch method \cite{rosenbrock1960automatic} or the
trust region algorithm \cite{byrd1987trust}, at the cost of computional
time. In this paper, we will restrict our attention to the performance
and pitfalls of Levenberg-Marquart, hence applying such methods is
beyond the scope of this paper.

\subsection{Fitting full log-periodic functions with AR(1) noise}

\begin{figure*}[!t]
\centerline{\includegraphics[width=0.4\textwidth]{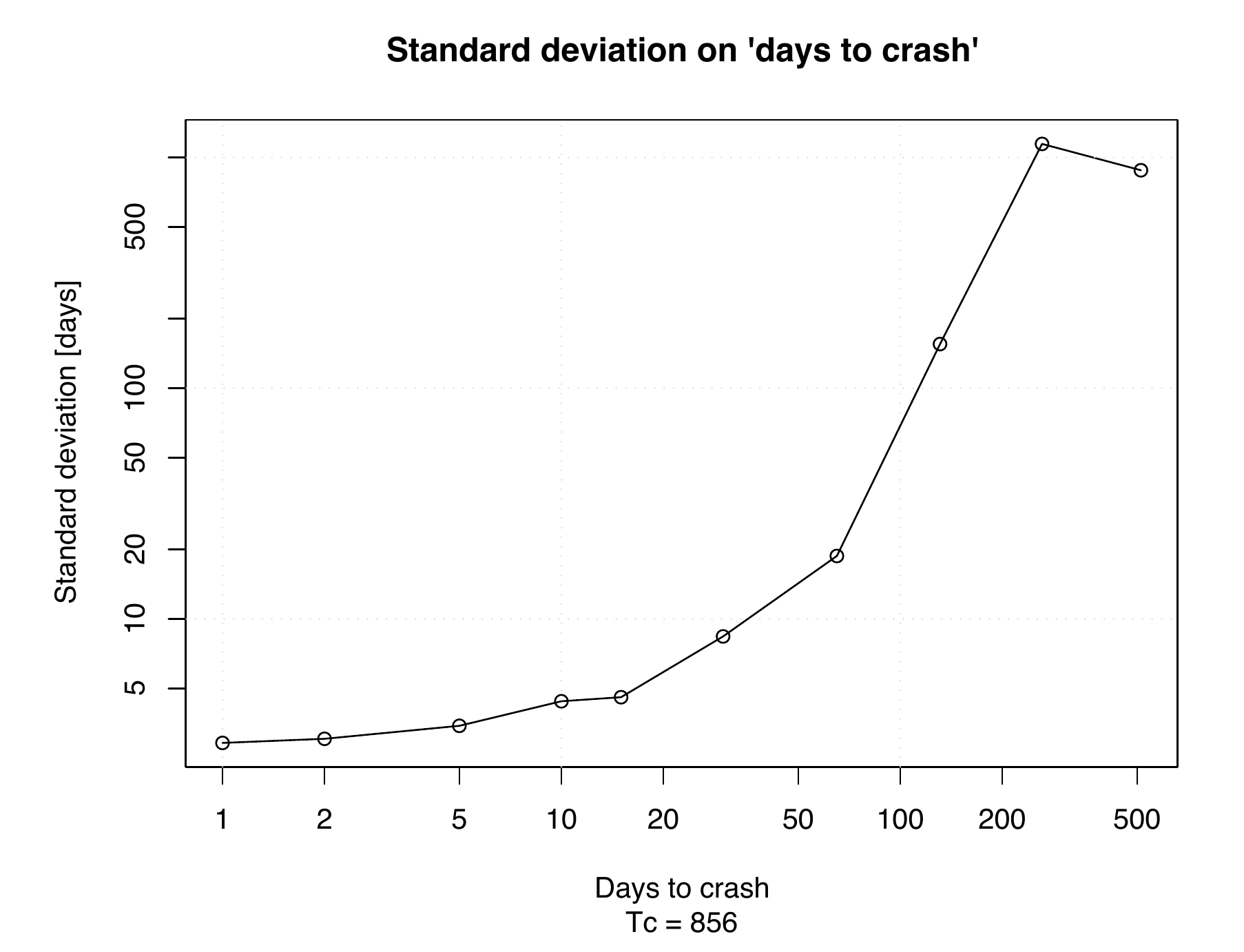}\includegraphics[width=0.4\textwidth]{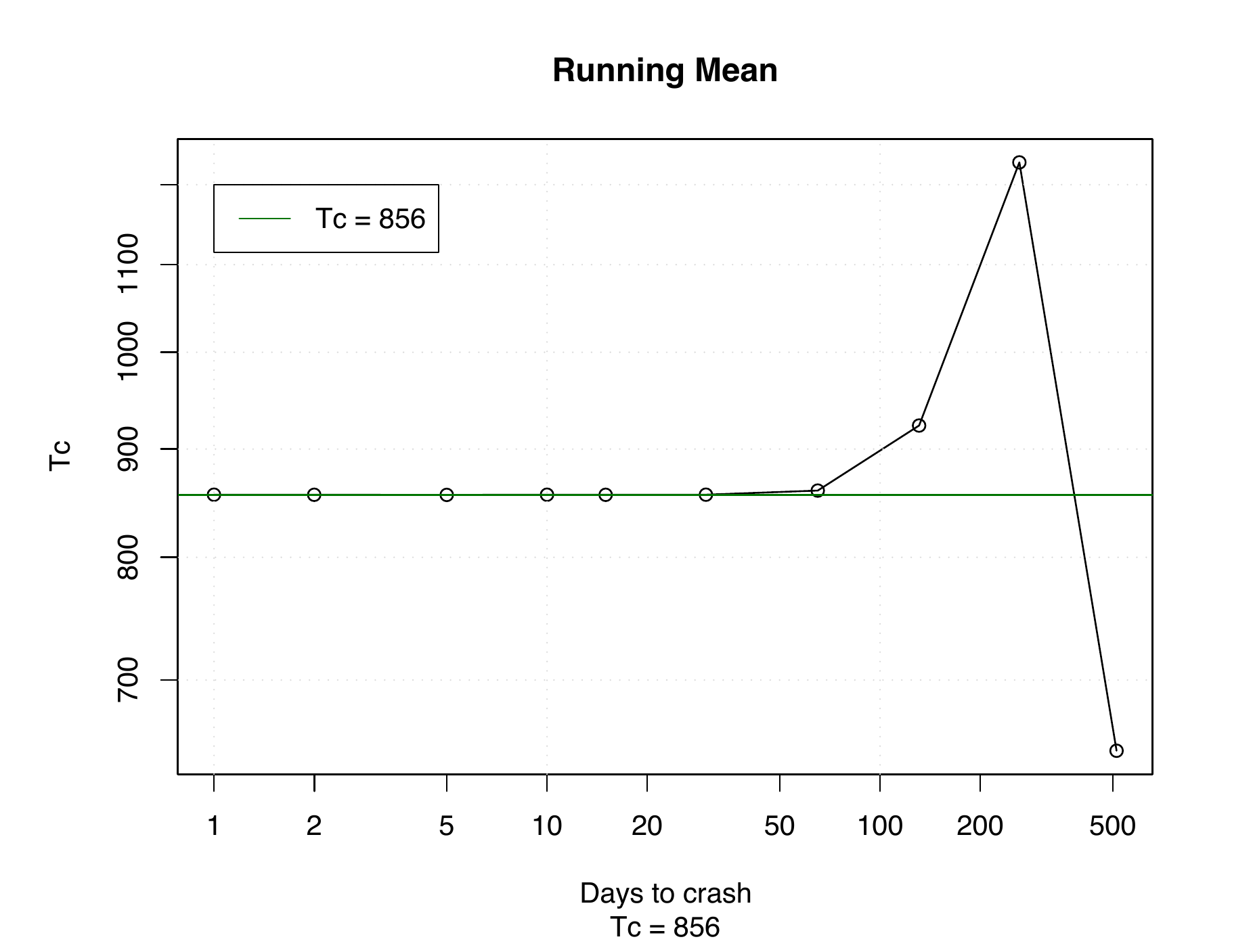}}
\caption{Average and standard deviation of the crash time estimate $\hat{{t_{c}}}$
for synthetic data with AR(1) noise and parameters reproducing the
1987 Hang-Sen crash.\label{fig:synth_data_avg_std}}
 
\end{figure*}

Among the recent progresses, the residuals were shown to be AR(1)
\cite{GFPR:2008,LRS:2009}. It makes sense, therefore, to create artificial
data with AR(1) noise. Let us consider the very simple case where
one adds some noise to a pure log-periodic function and applies a
fitting procedure. More specifically, we fit $f_{LP}(t)+\sigma\eta(t)$
where $\eta$ follows an auto-regressive process $\eta(t)=\eta(t-1)(1-\lambda)+\epsilon(t)$,
where $\epsilon\sim\mathcal{N}(0,1)$, $\lambda$ is the memory loss
and $\sigma$ tunes the strength of the fluctuations. AR(1) noise
that mimicks the fit of LPPL functions to the 1987 crash is obtained
with $\lambda=0.06$ and $\sigma=25$. A natural test of the predicting
power of the fit to $f_{LP}$ is to consider a time series that starts
at $t=1$ and keeps expanding until $t_{c}$. We created 1000 such
samples and computed averages of fitted parameters for increasing
time series length. The average estimates of the parameters, quite
remarkably including $t_{c}$, do converge to the true value at about
60 time steps (2.7 trading months) before the crash itself (Figure
\ref{fig:synth_data_avg_std}). Thus it turns out that fitting an
LPPL to synthetic data generated by an LPPL with a level of noise
comparable to that of real markets is possible and that the average
estimate of $t_{c}$ behaves very well ahead of $t_{c}$. Therefore,
one concludes that Levenberg-Marquart works well for estimating average
parameter values for synthetic data with many samples. Then a natural
crash warning is obtained when the average of $t_{c}$ stabilises.

However, when given a single run, predicting $t_{c}$ is much more
difficult: the standard deviation on $t_{c}$ is about a half of $t_{c}-t$.
Hence, since the residuals are Gaussian distributed (we have checked
that it is the case), the 80\% confidence window, as chosen in recent
papers on predictions with LPPL \cite{BCSWZ:2009,YWS10,JZSWBC:2010},
corresponds to a width of about $\frac{2}{3}(t_{c}-t)$, hence ranges
from $t_{c}-(t_{c}-t)/3$ to $t_{c}+(t_{c}-t)/3$, while the 95\%
confidence ranges from $t$ to $t_{c}+(t_{c}-t)$. So when a crash
warning is issued, the crash can occur any day at 95\% confidence.
Hence, predicting the date of a divergence is hard, even when the
underlying time series is a real LPPL. The 1987 crash was chosen because
LPPL fits it better than other ones. Hence, the above results yield
worse results for the parameters associated with other crashes.

\section{Discussion}

Given the attention devoted to LPPL and despite recent technical developments,
it is important to realise how sloppy this kind of function is. The
sloppiness of LPPLs implies that special care must be given when estimating
the uncertainty on $t_{c}$. The leap of faith of LPPL regarding bubbles
is not the log-periodic nature of oscillations, but to try to fit
data with functions that contain a divergence. Thus the discussion
on $t_{c}$ is largely disconnected with the nature of the oscillations,
as it is only related to a way to describe super-exponential growth.
Obviously one can fit real data with a function that does not contain
oscillations by setting $C=0$, thus focusing on the super-exponential
growth. We tried it on real data and while the precision on $t_{c}$
is slightly worse than that obtained with a LPPL, it is a simpler
method of obtaining an estimate for $t_{c}$.

%
{}

The fit of synthetic data with AR(1) noise is most revealing for several
reasons: it first shows that Levenberg-Marquart algorithm is adequate
for noisy \emph{synthetic} data, that is, when the underlying function
is of LPPL type. Next, the uncertainty associated with $t_{c}$ in
a realistic but nice case is quite large and is at the frontier of
being exploitable. This strongly suggests that making predictions
with real data is likely to yield worse uncertainties, since there
is a priori no reason for the oscillations of real data to be systematically
LPPL-based. Recent work that tries to estimate a probabilistic 80\%
confidence time window for $t_{c}$ is certainly a step in the right
direction. But since the time windows usually proposed are more optimistic
than the reference case considered here, it is very likely that the
method used underestimates the uncertainty on $t_{c}$. This is but
an example of the problem of estimating parameter uncertainty from
a single realisation of noise. As explained above, $t_{c}$ may fluctuate
very much when the time series given in input is changed slightly
because it is a sloppy parameter; hence, the mere fact that it does
fluctuate is not an indication per se that the variance of the fluctuations
approximates correctly its real uncertainty. Obtaining trustworthy
predictions for sloppy parameters from a single time series is possible
by Bayesian estimation \cite{brown2003statistical,frederiksen2004bayesian}.
Further work will look further in this direction.

\bibliographystyle{plain}
\bibliography{sloppylogper-2,biblio2}

\begin{thebibliography}{10}

\bibitem{Note1}
For example \cite {LRS:2009} recently claim that the LPPL model has few false
  positives on the grounds that all five of the bubbles that they identified on
  the S\&P500 between 3 January 1950 and 21 November 2008 {}``preceded
  well-know crashes.'' Using a criterion for a crash of a drop by > 15\% within
  12 weeks identifies 12 crashes in this period, but only two of these
  (25/8/1987, 17/7/1998) were among the five claimed by \cite [Table
  3]{LRS:2009}; the other three {}``well-known crashes'' (in 1994, 1997 and
  1999) do not meet these quite lenient crash criteria.

\bibitem{Note2}
While the beginning of a bubble is supposed to be given by the lowest value
  since the previous crash, it may be moved from this minimum to some later
  date; in \cite {JS:2001} the beginnings of half of the eight bubbles on the
  Hang Seng up to 1998 were thus moved \cite {BJ:2010}.

\bibitem{Note3}
Lomb periodograms \cite {B:2003} or the residuals \cite {GFPR:2008,LRS:2009}
  have been proposed, with mixed results.

\bibitem{BCSWZ:2009}
K.~Bastiaensen, P.~Cauwels, D.~Sornette, R.~Woodard, and W.-X. Zhou.
\newblock The \uppercase{C}hinese equity bubble: ready to burst.
\newblock http://www.scribd.com/doc/17337353/China-Bubble, July 2009.

\bibitem{B:1997}
F.~M. Borodich.
\newblock Renormalization schemes for earthquake prediction.
\newblock {\em Geophys Journal International}, 131(1):171--178, 1997.

\bibitem{BM:2003}
H.~C.~G. Bothmer and C.~Meister.
\newblock Predicting critical crashes? \uppercase{A} new restriction for the
  free variables.
\newblock {\em Physica A - Statistical Mechanics and its Applications},
  320:539--547, 2003.

\bibitem{BJ:2010}
D.~S. Br\'{e}e and N.L. Joseph.
\newblock Fitting the log periodic power law to financial crashes: a critical
  analysis.
\newblock arXiv:1002.1010, February 2010.

\bibitem{brown2003statistical}
K.S. Brown and J.P. Sethna.
\newblock {Statistical mechanical approaches to models with many poorly known
  parameters}.
\newblock {\em Physical Review E}, 68(2):21904, 2003.

\bibitem{byrd1987trust}
R.H. Byrd, R.B. Schnabel, and G.A. Shultz.
\newblock {A trust region algorithm for nonlinearly constrained optimization}.
\newblock {\em SIAM Journal on Numerical Analysis}, 24(5):1152--1170, 1987.

\bibitem{CF:2006}
G.~Chang and J.~Feigenbaum.
\newblock A \uppercase{B}ayesian analysis of log-periodic precursors to
  financial crashes.
\newblock {\em Quantitative Finance}, 6:15--36, 2006.

\bibitem{DKO:2008}
S.~Dro\.{z}d\.{z}, J.~Kwapie\'{n}, and P.~O\'{s}wie\c{c}imka.
\newblock Criticality charateristics of current oil price dynamics.
\newblock {\em Acta Physica Polonica A}, 114:699--702, 2008.

\bibitem{DKOS:2008}
S.~Dro\.{z}d\.{z}, J.~Kwapie\'{n}, P.~O\'{s}wie\c{c}imka, and J.~Speth.
\newblock Current log-periodic view on future world market development.
\newblock arXiv:0802.4043, 2008.

\bibitem{FFS:2009}
J.~Faillettaz, M.~Funk, and D.~Sornette.
\newblock Icequakes as precursors of ice avalanches.
\newblock arXiv:0906.5528, June 2009.

\bibitem{F:2001}
J.~Feigenbaum.
\newblock A statistical analysis of log-periodic precursors to financial
  crashes.
\newblock {\em Quantitative Finance}, 1:346--360, 2001.

\bibitem{frederiksen2004bayesian}
S.L. Frederiksen, K.W. Jacobsen, K.S. Brown, and J.P. Sethna.
\newblock {Bayesian ensemble approach to error estimation of interatomic
  potentials}.
\newblock {\em Physical review letters}, 93(16):165501, 2004.

\bibitem{GFPR:2008}
L.~Gazola, C.~Fernades, A~Pizzinga, and R.~Riera.
\newblock The log-periodic-{AR}(1)-{GARCH}(1,1) model of financial crashes.
\newblock {\em European Physics Journal B}, 61:355--362, 2008.

\bibitem{gutenkunst2007universally}
R.N. Gutenkunst, J.J. Waterfall, F.P. Casey, K.S. Brown, C.R. Myers, and J.P.
  Sethna.
\newblock {Universally sloppy parameter sensitivities in systems biology
  models}.
\newblock {\em PLoS Comput Biol}, 3(10):1871--1878, 2007.

\bibitem{HSS:2000}
Y.~Huang, H~. Saleur, and D.~Sornette.
\newblock Reexamination of log periodicity observed in the seismic precursors
  of the 1989 {L}oma {P}rieta earthquake.
\newblock {\em Journal of Geophysical Research}, 105(B12):28111--28123, 2000.

\bibitem{J:2009}
E.~Jacobsen.
\newblock How to predict crashes in financial markets with the log-periodic
  power law.
\newblock Master's thesis, Mathematical Statistics, Stockholm University, 2009.

\bibitem{JZSWBC:2010}
Z.-Q. Jiang, W.-X. Zhou, D.~Sornette, R.~Woodard, K.~Bastiaensen, and
  P.~Cauwels.
\newblock Bubble diagnosis and prediction of the 2005-2007 and 2008-2009
  {C}hinese stock market bubbles.
\newblock {\em Journal of Economic Behavior and Organization}, in press.

\bibitem{JLS:2000}
A.~Johansen, O.~Ledoit, and D.~Sornette.
\newblock Crashes as critical points.
\newblock {\em International Journal of Theoretical and Applied Finance},
  3:219--225, 2000.

\bibitem{JS:2001}
A.~Johansen and D.~Sornette.
\newblock Bubbles and anti-bubbles in \uppercase{L}atin-\uppercase{A}merican,
  \uppercase{A}sian and \uppercase{W}estern stock markets: an empirical study.
\newblock {\em International Journal of Theoretical and Applied Finance},
  4:853--920, 2001.

\bibitem{CKYOSTS:2008}
Y.~Kawada, H.~Nagahama, Y.~Yasuoka, Y.~Omori, T.~Ishikawa, S.~Tokonami, and
  M.~Shinogi.
\newblock Power-law and log-periodic changes in precursory phenomena prior to
  large earthquakes.
\newblock In {\em The 33rd International Geological Congress}, Oslo, August
  2008.

\bibitem{LPC:1999}
L.~Laloux, M.~Potters, and R.~Cont.
\newblock Are financial crashes predictable?
\newblock {\em Europhysics Letters}, 45:1--5, 1999.

\bibitem{LM:2004}
F.~Lillo and R.~Mantegna.
\newblock Dynamics of a financial market index after a crash.
\newblock {\em Physica A}, 338:125--134, 2004.

\bibitem{LRS:2009}
L.~Lin, R.E. Ren, and D.~Sornette.
\newblock A consistent model of `explorative' financial bubbles with
  mean-reversing residuals.
\newblock arXiv:0905.0128v1, May 2009.

\bibitem{MMT:2005}
B.~D. Malamud, G.~Morein, and D.~L. Turcotte.
\newblock Log-periodic behavior in a forest-fire model.
\newblock {\em Nonlinear Processes in Geophysics}, 12:575--585, 2005.

\bibitem{MGFD:2005}
R.~Matsushita, I.~Gleria, A.~Figueiredo, and S.~Da Silva.
\newblock On log-periodic crashes.
\newblock Finance 0505007, Economics Working Paper Archive, May 2005.

\bibitem{NTG:1995}
W.~I. Newman, D.~L. Turcotte, and A.~M. Gabrielov.
\newblock Log-periodic behavior of a hierarchical failure model with
  applications to precursory seismic activation.
\newblock {\em Physical Review E}, 52:4827--4835, 1995.

\bibitem{NCG:2000}
L.~Nottale, J.~Chaline, and P.~Grou.
\newblock On the fractal structure of evolutionary trees.
\newblock In G.~Losa, D.~Merlini, T.~Nonnenmacher, and E.~Weibel, editors, {\em
  Fractals in Biology and Medicine}, volume~3, pages 247--258, Ascona,
  Switzerland, March 2000. Birkh{\"a}user Verlag.

\bibitem{rosenbrock1960automatic}
H.~H. Rosenbrock.
\newblock {An automatic method for finding the greatest or least value of a
  function}.
\newblock {\em The Computer Journal}, 3(3):175, 1960.

\bibitem{SJ:1997}
D.~Sornette and A.~Johansen.
\newblock Large financial crashes.
\newblock {\em Physica A}, 245:411--422, 1997.

\bibitem{SJ:2001}
D.~Sornette and A.~Johansen.
\newblock Significance of log-periodic precursors to financial crashes.
\newblock {\em Quantitative Finance}, 1:452--471, 2001.

\bibitem{sornette1996stock}
D.~Sornette, A.~Johansen, and J.-P. Bouchaud.
\newblock Stock market crashes, precursors and replicas.
\newblock {\em Journal of Physics I France}, 6:167--175, 1996.

\bibitem{SZ:2006}
D.~Sornette and W.-X. Zhou.
\newblock Predictability of large future changes in major financial indices.
\newblock {\em International Journal of Forecasting}, 22:153--168, 2006.

\bibitem{SH:2005}
K.~Suchecki and J.~A. Holyst.
\newblock Log-periodic oscillations in degree distributions of hierarchical
  scale-free networks.
\newblock {\em Acta Physica Polonica B}, 36(8):2499--2511, 2005.

\bibitem{T:2002}
H.~Tanaka.
\newblock Log-periodic modulation to power law of change in the geoelectric
  potential difference observed on {N}iijima {I}sland prior to volcanic and
  seismic activation in the {I}zu {I}sland region, {J}apan 2000.
\newblock {\em Proceedings of the Japanese Academy Series B}, 78B(9):271--276,
  2002.

\bibitem{B:2003}
H.~G. van Bothmer.
\newblock Significance of log-periodic signatures in cumulative noise.
\newblock {\em Quantitative Finance}, 3:370--375, 2003.

\bibitem{BosSumExp}
A.~van~den Bos and J.~H. Swarte.
\newblock Resolvability of the parameters of multiexponentials and other sum
  models.
\newblock {\em IEEE Transactions on Signal Processing}, 41(1):313, 1993.

\bibitem{SethnaSloppyPRL}
J.~J. Water, F.~P. Casey, R.~N. Gutenjunst, K.~S. Brown, C.~R. Myers, P.~W.
  Brouwer, V.~Elser, and J.~P. Sethna.
\newblock Sloppy-model universality class and the {V}andermond matrix.
\newblock {\em Phys. Rev. Lett.}, 97:150601, 2006.

\bibitem{YWS10}
W.~Yana, R.~Woodard, and D.~Sornette.
\newblock Diagnosis and prediction of tipping points in financial markets:
  crashes and rebounds.
\newblock arXiv:1003.5926v1, March 2010.

\bibitem{ZS:2006}
W.~Zhou and D.~Sornette.
\newblock Is there a real-estate bubble in the us?
\newblock {\em Physica A: Statistical Mechanics and its Applications},
  361(1):297--308, 2006.

\end{thebibliography}

\end{document}